\documentclass[useAMS,usenatbib]{mn2e}
\usepackage{multirow}
\usepackage{epsfig}
\usepackage{subfigure}
\usepackage{url}

\makeatletter
\let\jnl@style=\rm
\def\ref@jnl#1{{\jnl@style#1}}

\def\aj{\ref@jnl{AJ}}                   
\def\araa{\ref@jnl{ARA\&A}}             
\def\apj{\ref@jnl{ApJ}}                 
\def\apjl{\ref@jnl{ApJ}}                
\def\apjs{\ref@jnl{ApJS}}               
\def\ao{\ref@jnl{Appl.~Opt.}}           
\def\apss{\ref@jnl{Ap\&SS}}             
\def\aap{\ref@jnl{A\&A}}                
\def\aapr{\ref@jnl{A\&A~Rev.}}          
\def\aaps{\ref@jnl{A\&AS}}              
\def\azh{\ref@jnl{AZh}}                 
\def\baas{\ref@jnl{BAAS}}               
\def\jrasc{\ref@jnl{JRASC}}             
\def\memras{\ref@jnl{MmRAS}}            
\def\mnras{\ref@jnl{MNRAS}}             
\def\pra{\ref@jnl{Phys.~Rev.~A}}        
\def\prb{\ref@jnl{Phys.~Rev.~B}}        
\def\prc{\ref@jnl{Phys.~Rev.~C}}        
\def\prd{\ref@jnl{Phys.~Rev.~D}}        
\def\pre{\ref@jnl{Phys.~Rev.~E}}        
\def\prl{\ref@jnl{Phys.~Rev.~Lett.}}    
\def\pasp{\ref@jnl{PASP}}               
\def\pasj{\ref@jnl{PASJ}}               
\def\qjras{\ref@jnl{QJRAS}}             
\def\skytel{\ref@jnl{S\&T}}             
\def\solphys{\ref@jnl{Sol.~Phys.}}      
\def\sovast{\ref@jnl{Soviet~Ast.}}      
\def\ssr{\ref@jnl{Space~Sci.~Rev.}}     
\def\zap{\ref@jnl{ZAp}}                 
\def\nat{\ref@jnl{Nature}}              
\def\iaucirc{\ref@jnl{IAU~Circ.}}       
\def\aplett{\ref@jnl{Astrophys.~Lett.}} 
\def\apspr{\ref@jnl{Astrophys.~Space~Phys.~Res.}}
\def\bain{\ref@jnl{Bull.~Astron.~Inst.~Netherlands}}
\def\fcp{\ref@jnl{Fund.~Cosmic~Phys.}}  
\def\gca{\ref@jnl{Geochim.~Cosmochim.~Acta}}   
\def\grl{\ref@jnl{Geophys.~Res.~Lett.}} 
\def\jcp{\ref@jnl{J.~Chem.~Phys.}}      
\def\jgr{\ref@jnl{J.~Geophys.~Res.}}    
\def\jqsrt{\ref@jnl{J.~Quant.~Spec.~Radiat.~Transf.}}
\def\memsai{\ref@jnl{Mem.~Soc.~Astron.~Italiana}}
\def\nphysa{\ref@jnl{Nucl.~Phys.~A}}   
\def\physrep{\ref@jnl{Phys.~Rep.}}   
\def\physscr{\ref@jnl{Phys.~Scr}}   
\def\planss{\ref@jnl{Planet.~Space~Sci.}}   
\def\procspie{\ref@jnl{Proc.~SPIE}}   

\makeatother

\newcommand {\apgt} {\ {\raise-.5ex\hbox{$\buildrel>\over\sim$}}\ }
\newcommand {\aplt} {\ {\raise-.5ex\hbox{$\buildrel<\over\sim$}}\ } 

\title[Unveiling the nucleus of NGC 1068]{{\it NuSTAR} catches the unveiling nucleus of NGC 1068}

\author[Andrea Marinucci, et al.]{A. Marinucci$^{1}$\thanks{E-mail: marinucci@fis.uniroma3.it (AM)}, S. Bianchi$^{1}$, G. Matt$^{1}$, D. M. Alexander$^{2}$, M. Balokovi\'{c}$^{3}$, 
\newauthor
F. E. Bauer$^{4,5,6}$, W. N. Brandt$^{7,8,9}$, P. Gandhi$^{2}$, M.Guainazzi$^{10}$, F. A. Harrison$^{3}$, 
\newauthor
K. Iwasawa$^{11}$, M. Koss$^{12}$, K. K. Madsen$^{3}$, F. Nicastro$^{13,14,15}$, S. Puccetti$^{13,16}$, 
\newauthor
C. Ricci$^{4}$, D. Stern$^{17}$, D. J. Walton$^{17,3}$ \\
$^1$Dipartimento di Matematica e Fisica, Universit\`a degli Studi Roma Tre, via della Vasca Navale 84, 00146 Roma, Italy\\
$^{2}$Department of Physics, Durham University, Durham DH1 3LE, UK\\
$^3$Cahill Center for Astronomy and Astrophysics, California Institute of Technology, Pasadena, CA, 91125 USA\\
$^4$Instituto de Astrof\'{i}sica, Facultad de F\'{i}sica, Pontificia Universidad Cat\'{o}lica de Chile, 306, Santiago 22, Chile\\
$^5$Millennium Institute of Astrophysics, Vicu\~{n}a Mackenna 4860, 7820436 Macul, Santiago, Chile\\
$^6$Space Science Institute, 4750 Walnut Street, Suite 205, Boulder, Colorado 80301\\
$^{7}$Department of Astronomy \& Astrophysics, 525 Davey Lab, The Pennsylvania State University, University Park, PA 16802, USA\\
$^{8}$Institute for Gravitation and the Cosmos, The Pennsylvania State University, University Park, PA 16802, USA\\
$^{9}$Department of Physics, 104 Davey Lab, The Pennsylvania State University, University Park, PA 16802, USA\\
$^{10}$European Space Astronomy Centre of ESA, PO Box 78, Villanueva de la Canada, 28691, Madrid, Spain\\
$^{11}$ICREA and Institut de Ci\`encies del Cosmos, Universitat de Barcelona, IEEC-UB, Mart\'i i Franqu\`es, 1, 08028 Barcelona, Spain\\
$^{12}$Institute for Astronomy, Department of Physics, ETH Zurich, Wolfgang-Pauli-Strasse 27, CH-8093 Zurich, Switzerland\\
$^{13}$INAF Osservatorio Astronomico di Roma, via Frascati 33,00040 Monte Porzio Catone (RM), Italy\\
$^{14}$Harvard-Smithsonian Center for Astrophysics, 60 Garden Street, MS-04, Cambridge, MA 02138, USA\\
$^{15}$Department of Physics, University of Crete, PO Box 2208, GR-710 03 Heraklion, Crete, Greece \\
$^{16}$ASDC-ASI, Via del Politecnico, 00133 Roma, Italy\\
$^{17}$Jet Propulsion Laboratory, California Institute of Technology, 4800 Oak Grove Drive, Pasadena, CA 91109, USA\\
}

\begin{document}
\maketitle
\label{firstpage}

\begin{abstract} 
We present a {\it NuSTAR} and XMM-{\it Newton} monitoring campaign in 2014/2015 of the Compton-thick Seyfert 2 galaxy, NGC 1068. During the August 2014 observation, we detect with {\it NuSTAR} a flux excess above 20 keV ($32\pm6 \%$) with respect to the December 2012 observation and to a later observation performed in February 2015. We do not detect any spectral variation below 10 keV in the XMM-{\it Newton} data. The transient excess can be explained by a temporary decrease of the column density of the obscuring material along the line of sight (from N$_{\rm H}\simeq10^{25}$ cm$^{-2}$ to N$_{\rm H}=6.7\pm1.0\times10^{24}$ cm$^{-2}$), which allows us for the first time to unveil the direct nuclear radiation of the buried AGN in NGC 1068 and to infer an intrinsic 2--10 keV luminosity L$_{\rm X}=7^{+7}_{-4} \times 10^{43}$ erg s$^{-1}$.
\end{abstract}

\begin{keywords}
Galaxies: active - Galaxies: Seyfert - Galaxies: accretion - Individual: NGC 1068
\end{keywords}

\section{Introduction}

Since \citet{am85} proposed the unification scheme for type-1 and type-2 AGN, it has been commonly thought that highly absorbed (i.e. Compton-thick, with N$_{\rm H}\geq 1.5 \times 10^{24}$ cm$^{-2}$) Seyfert 2s are obscured by neutral gaseous matter embedded in a thick molecular torus located at parsec distances from the central X-ray source \citep[see][for a recent review]{n15}.
Reflection from the torus reveals itself through a very intense neutral Iron K$\alpha$ emission line at 6.4 keV, with equivalent widths of $\sim 1$ keV, and a prominent Compton hump peaking at $\sim20$ keV \citep{ghm94}.

Recently, both the size and the distance of this thick screen have been questioned by a number of observations that have
measured significant column density variability of the innermost absorber over time-scales of days or even
hours in nearby bright sources such as NGC 1365 \citep{ris05, rrw15}, NGC 4388 \citep{elvis04}, NGC~4151 \citep{puc07} and NGC 7582 \citep[][Rivers et al., in prep.]{bianchi09c}. On the other hand, spatially resolved Iron K$\alpha$ line emission, extended on scales of hundreds of parsecs, has been detected in the brightest Compton-thick objects such as NGC 1068 \citep{yws01,brink02}, NGC 4945 \citep{mrw12} and Mrk~3 \citep{glm12}. These measurements suggest that different absorbers/reflectors, located on a variety of spatial scales, may contribute \citep[see e.g. ][]{bmr12} to the absorption and reprocessing. 

NGC 1068 \citep[$\rm D_L$=14.4 Mpc:][]{t88} is one of the best studied Seyfert 2 galaxies. Indeed, the unification model was first proposed to explain the presence of broad optical lines in its polarized light.
In X-rays, it was first studied by {\it Ginga}, which detected a strong (EW$\sim$1.3 keV) neutral Iron line \citep{koy89}, an unambiguous sign that we are observing reflected, rather than direct, radiation \citep{mbf96}. This result was later confirmed by {\it ASCA} \citep{ueno94,ifm97} which resolved the Iron line into neutral and ionized components.
{\it BeppoSAX} \citep{matt97} found no evidence for transmitted radiation up to 100
keV, implying a column density of the absorbing material in excess of 10$^{25}$ cm$^{-2}$. \citet{matt04} and \citet{pv06} studied the XMM-{\it Newton}/EPIC spectra, and found evidence for an iron overabundance with respect to the solar value.

Recently, \citet{baw15} analysed the multi-epoch X-ray spectra of NGC 1068 using different observatories, including 3-79 keV data from {\it NuSTAR}. They interpreted the broadband cold reflected emission of NGC 1068 as originating from multiple reflectors with three distinct column densities.  The highest N$_{\rm H}$ component (N$_{\rm H,1}\simeq10^{25}$ cm$^{-2}$) is the dominant contribution to the Compton hump, while the lowest N$_{\rm H}$ component (N$_{\rm H,2}\sim1.5\times10^{23}$ cm$^{-2}$) produces much of the line emission. The authors also confirm that almost 30\% of the neutral Fe K$\alpha$ line flux arises from regions outside the central $140$ pc.

\citet{gmv00} found evidence for variability, in the 3--10 keV band, comparing two {\it BeppoSAX} observations performed about one year apart.
Later on, comparing {\it ASCA}, {\it RossiXTE} and {\it BeppoSAX} spectra taken at different epochs spanning a few months, \citet{cwk02} claimed variations in both the continuum and He-like iron line flux on time scales as short as four months, using the 2--10 keV energy band. 
\citet{matt04}, comparing an XMM-{\it Newton} observation with {\it BeppoSAX} observations performed a few years earlier, found possible evidence for
flux variability of both the cold and the ionized reflectors.

We present a joint XMM-{\it Newton} and {\it NuSTAR} monitoring campaign of NGC 1068, from July 2014 until February 2015, and report on the discovery of a transient excess above 20 keV. 

We adopt the cosmological parameters $H_0=70$ km s$^{-1}$ Mpc$^{-1}$, $\Omega_\Lambda=0.73$ and $\Omega_m=0.27$, i.e. the default ones in \textsc{xspec 12.8.1} \citep{xspec}. Errors correspond to the 90\% confidence level for one interesting parameter ($\Delta\chi^2=2.7$), if not stated otherwise. 

\begin{figure}
\epsfig{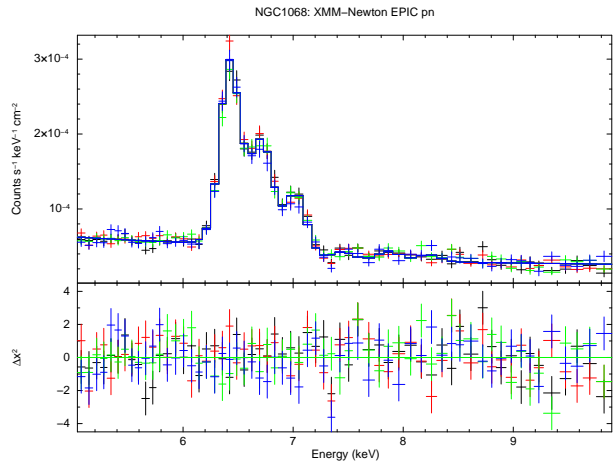}
\caption{\label{xmm_plot} Best fit of the four EPIC-pn spectra, with residuals. Black, red, green and blue data points indicate observations performed on 2014 July 10, July 18, August 19 and 2015 February 3, respectively.}
\end{figure}

\section{Observations and data reduction}
{\bf NuSTAR}. NGC 1068 was observed by {\it NuSTAR} with its two coaligned X-ray telescopes five times. The first three times were in December 2012: those data are discussed in \citet{baw15}. After that, NGC 1068 was the target of a monitoring campaign with XMM-{\it Newton} composed of four observations, from July 2014 until February 2015. {\it NuSTAR} observed the source simultaneously with the third and fourth XMM-{\it Newton} pointings. The Level 1 data products were processed with the {\it NuSTAR} Data Analysis Software (NuSTARDAS) package (v. 1.3.0). Cleaned event files (level 2 data products) were produced and calibrated using standard filtering criteria with the \textsc{nupipeline} task and the latest calibration files available in the {\it NuSTAR} calibration database (CALDB 20150316). Since no spectral variation is found within each observation, we decided to use time-averaged spectra for each epoch. Background light curves are constant within each observation and do not present any flares due to spurious emission. The background levels are perfectly consistent between the three epochs. We coadded data taken in December 2012, since no variation was found between those three pointings. The extraction radii for the source and background spectra were $1.5$ arcmin each. Net exposure times, after this process, can be found in Table \ref{nuxmmobs}, for both Focal Plane Modules A and B. 
The two pairs of {\it NuSTAR} spectra were binned in order to over-sample the instrumental resolution by at least a factor of 2.5 and to have a Signal-to-Noise Ratio (SNR) greater than 5 in each spectral channel.\\
{\bf XMM-Newton}. The monitoring campaign of NGC 1068 with XMM-{\it Newton} was composed of four $\sim40$ ks observations, starting on 2014 July 10  with the EPIC CCD cameras, the pn \citep{struder01} and the two MOS \citep{turner01}, operated in small window and thin filter mode. Details of the  XMM-{\it Newton} observations and analysis can be found in Bianchi et al. (in prep.).  
The resulting net exposure times can be found in Table \ref{nuxmmobs} for the EPIC-pn. Spectra were binned in order to over-sample the instrumental resolution by at least a factor of three and to have no less than 30 counts in each background-subtracted spectral channel.  Cross-calibration constants between the {\it NuSTAR}-FPMA/B and EPIC-pn are within 10\%, in agreement with values presented in \citet{mhm15}. 

NGC 1068 hosts a strong ULX near its nucleus already studied in \citet{matt04}. The source (at a distance of $\sim 28''$ from the AGN) is present in our 2014/2015 XMM data and we find no differences with respect to the properties discussed in \citet{matt04}: its contribution to the 4-10 keV spectrum is constrained to be $\leq 5\%$. 

\begin{figure*} 
\begin{center}
\epsfig{file=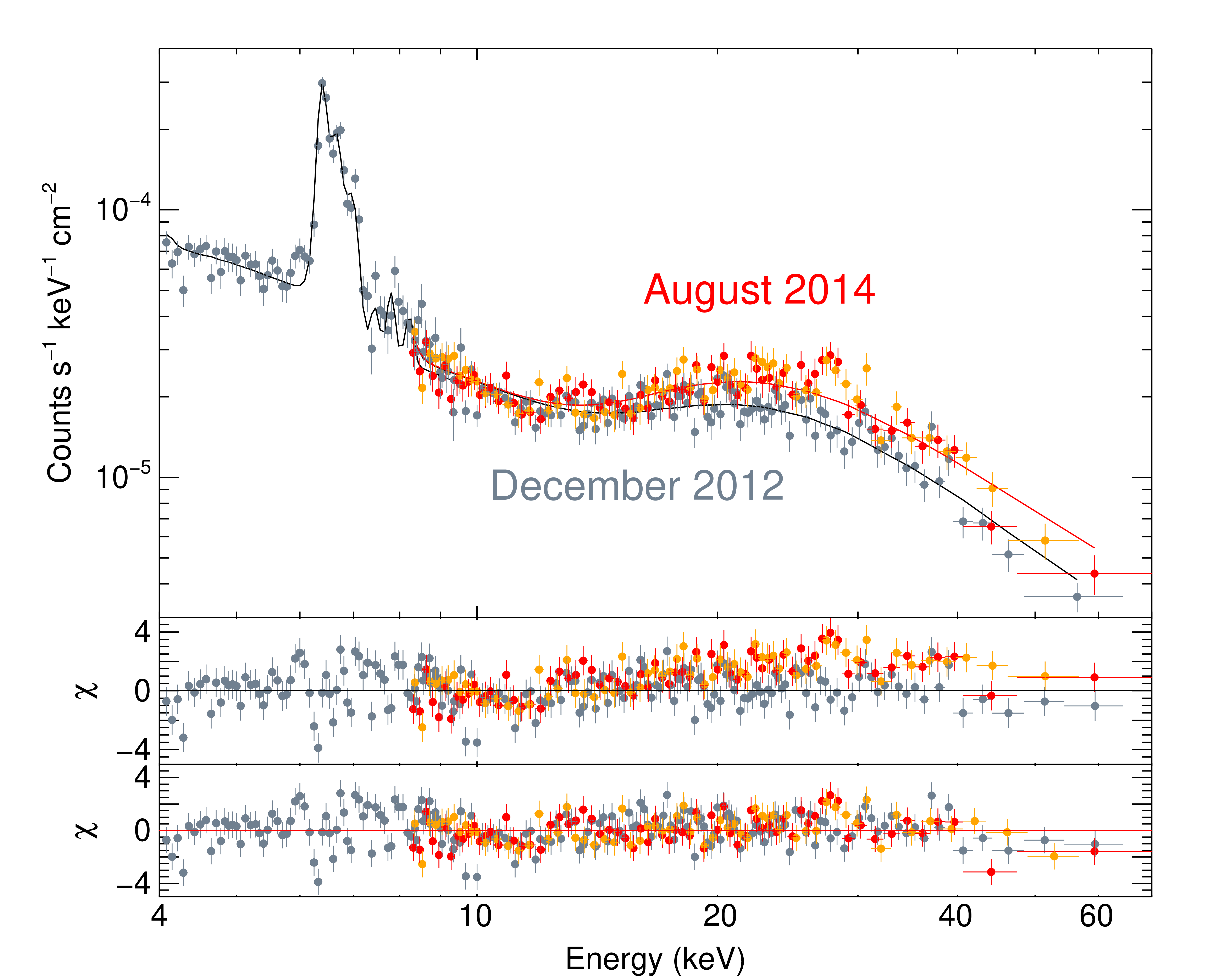, width=\columnwidth}
\epsfig{file=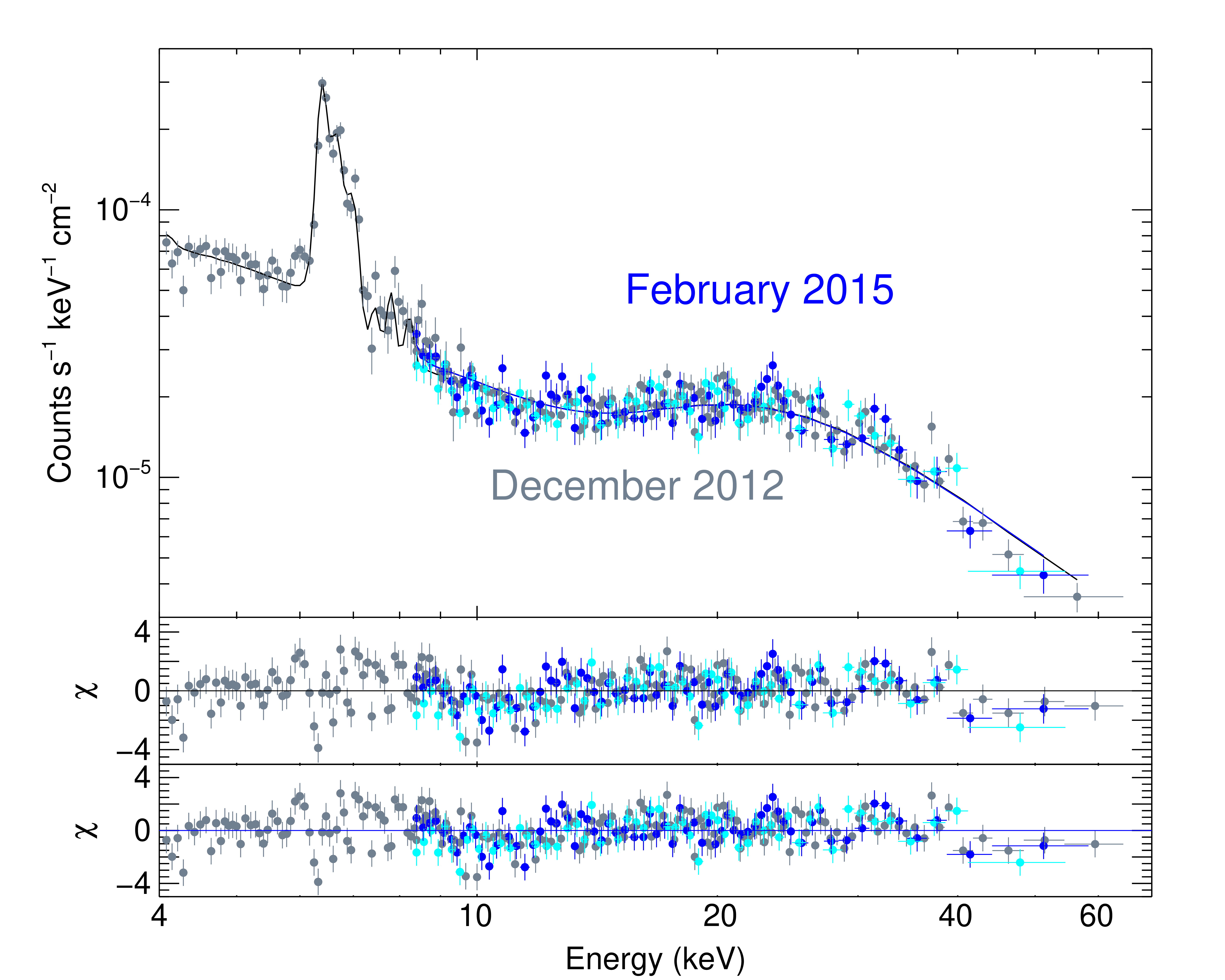, width=\columnwidth}
\caption{\label{PLOT} Best fit models, spectra and residuals are shown. The August 2014 EPIC-pn spectrum and the 2012 FPMA/B spectra are plotted as grey circles. The 2014 FPMA and FPMB data are overplotted in the top-left panel in red and orange, respectively. The new 2015 FPMA and FPMB data are overplotted in the top-right panel in blue and cyan, respectively. Middle panels show residuals of the two 2014 and 2015 data set to the 2012 best fit model (black solid line). Bottom panels show residuals to a model in which the column density and the nuclear flux are left free in the new observations (red and blue solid lines). See text for details.}
\end{center}
\end{figure*}

\begin{table*}
\begin{center}
\begin{tabular}{ccccccc}
{\bf Obs. ID} & {\bf Date} & {\bf Exp. Time (ks)} & \multicolumn{2}{c}{\bf 4--20 keV Count rate (cts s$^{-1}$)}&\multicolumn{2}{c}{\bf 20--80 keV Count rate (cts s$^{-1}$)}\\
\hline
 \multicolumn{1}{c} { } & & & {FPMA} & FPMB & FPMA & FPMB\\ 
60002030002 &2012-12-18 &57\multirow{3}{*}{\Bigg{\}}} & &\\
60002030004 &2012-12-20 & 48$\ \ \ $ &$0.1579\pm0.0011$ &$0.1494\pm0.0011$ & $0.0324\pm0.0006$&  $0.0286\pm0.0006$\\
60002030006 &2012-12-21 & 19$\ \ \ $& &\\
60002033002 &2014-08-18 & 52&$0.1556\pm0.0018$ &$0.1456\pm0.0018$& $0.0417\pm0.0010$& $0.0387\pm0.0010$\\
60002033004 &2015-02-05 & 53&$0.1573\pm0.0018$ &$0.1444\pm0.0017$& $0.0329\pm0.0009$& $0.0285\pm0.0008$\\
 &  &  & \multicolumn{2}{c}{\bf 4--10 keV Count rate (cts s$^{-1}$)}&\\
\hline
& & & \multicolumn{2}{c}{EPIC-pn}   & \\ 
0740060201&2014-07-10&44 &\multicolumn{2}{c}{$0.1593\pm0.0022$ }&\\
0740060301&2014-07-18& 39&\multicolumn{2}{c} {$0.1540\pm0.0021$}&\\
0740060401&2014-08-19& 37& \multicolumn{2}{c}{$0.1537\pm0.0021$}&\\
0740060501&2015-02-03&37 &\multicolumn{2}{c}{$0.1625\pm0.0025$} &\\
\hline
\end{tabular}
\end{center}
\caption{\label{nuxmmobs} Observation log for the {\it NuSTAR} and XMM-{\it Newton} monitoring of NGC 1068.}
\end{table*}

\section{Spectral analysis}
We start our analysis by checking for variability in the four XMM-{\it Newton} spectra obtained between July 2014 and February 2015. In our analysis, we only consider data above 4 keV due to the strong contribution at lower energies from distant photoionized, extra-nuclear emission, which will be discussed in  Bianchi et al. (in prep.).  No differences are found between the spectra, nor do we find differences between the new data set and the one taken in July 2000: we refer to Bianchi et al. (in prep.) for details. Applying the model discussed in \citet{matt04} for the old XMM observation to the new 5--10 keV pn spectra, we infer that the flux of the narrow core of the iron K$\alpha$ line is constant within 5\%. 
The best fit of the four spectra is shown in Fig. \ref{xmm_plot}: no spectral or flux variations are apparent.  \\
Since no significant changes are found between the XMM spectra, we used only the observation simultaneous to the high-energy transient event (ObsId 0740060401: Table 1, Fig. 2) and then included {\it NuSTAR} FPMA and FPMB spectra from observations performed in December 2012, August 2014 and February 2015. The final data set is therefore comprised of seven spectra: one XMM spectrum taken in August 2014 and three pairs of {\it NuSTAR} spectra taken in 2012, 2014 and 2015. We only considered {\it NuSTAR} data above 8 keV because XMM spectra have higher spectral resolution and higher SNR in the Fe K$\alpha$ energy range.

We then apply the best-fit model discussed in \citet{baw15} to our XMM+{\it NuSTAR} 4-79 keV data set. This model fits data from 1996 until 2012 from multiple X-ray observatories. The authors found, using \textsc{MYTorus} tables (model M2d), that the reflecting material is composed of three distinct components with N$_{\rm H,1}\simeq10^{25}$ cm$^{-2}$, N$_{\rm H,2}=(1.5\pm0.1)\times10^{23}$ cm$^{-2}$ and N$_{\rm H,3}=(5.0_{-1.9}^{+4.5})\times10^{24}$ cm$^{-2}$. N$_{\rm H,1}$ is the absorbing column density along the line of sight. {\it Chandra} observations show that the spectral features attributed to the N$_{\rm H,1}$ and N$_{\rm H,2}$ components arise from the central 2 arcsec only, while N$_{\rm H,3}$ corresponds to regions outside the central 2 arcsec. N$_{\rm H,1}$ and N$_{\rm H,3}$ contribute primarily to the Compton hump, while N$_{\rm H,2}$ and N$_{\rm H,3}$ provide dominant contributions to the Fe K lines \citep{baw15}. 

For this analysis the normalizations of the absorbed, scattered and line emission in \textsc{MYTorus} tables were kept tied together (coupled reprocessor solution; Yaqoob et al. 2012): further details about this solution can be found in \citet{baw15}. Applying their best-fit model to our data set we find $\chi^2$/dof=1166/896=1.30 because the model does not reproduce well the August 2014 {\it NuSTAR} data (Fig. \ref{PLOT}, left-middle panel). Indeed, Fig. \ref{PLOT} and Table \ref{nuxmmobs} show that there is a $32\pm6\%$ flux increase above 20 keV and a clear additional spectral feature in the {\it NuSTAR} observation taken in August 2014. 

\begin{figure} 
\begin{center}
\epsfig{file=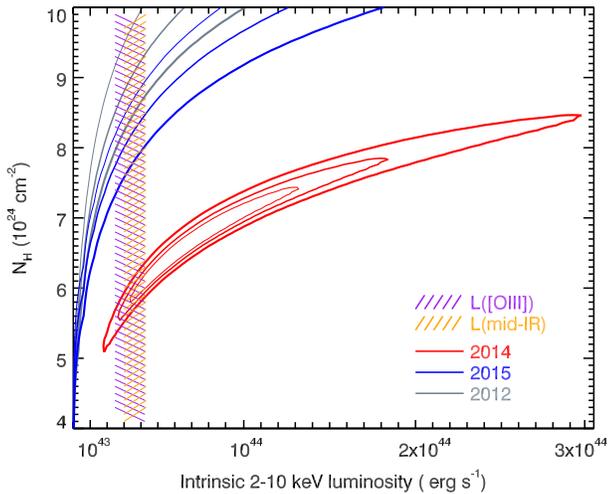, width=\columnwidth}
\caption{\label{cp} Contour plots between the column density along the line of sight and the intrinsic 2-10 keV luminosity. Grey, red and blue lines represent contours for 2012, 2014 and 2015 data sets, respectively. Boldest to thinnest contours indicate 99\%, 90\% and 68\% confidence levels. Vertical orange and purple stripes indicate the 2-10 keV intrinsic luminosities inferred from the mid-IR and [OIII] observed luminosities with their relative dispersions from \citet{ghs09} and \citet{lamastra09}. }
\end{center}
\end{figure}

We first model this excess by leaving the column density N$_{\rm H,1}$ (along the line of sight) free to vary in the 2014 spectra, which results in a significant improvement of the fit ($\Delta\chi^2$=177 for one additional free parameter). A marginal improvement is found when we also leave the normalization of the primary component free to vary ($\Delta\chi^2$=10 for one additional free parameter), for a final $\chi^2$/dof=979/894=1.09; no strong residuals are seen throughout the whole 4-79 keV energy band (Fig. 2, bottom panels). Best fit values for the column density along the line of sight and for the nuclear component normalization are N$_{\rm H,1}=(6.7\pm1.0)\times 10^{24}$ cm$^{-2}$ and A$_{\rm nucl}=0.9^{+1.0}_{-0.5}$ ph cm$^{-2}$ s$^{-1}$ keV$^{-1}$ at 1 keV, respectively.
This normalization leads to an unabsorbed 2--10 keV luminosity L$_{\rm X}=7^{+7}_{-4} \times 10^{43}$ erg s$^{-1}$ , which is consistent with the value presented in Bauer et al. (2015), within the error bars (Fig. 3).The intrinsic luminosity presented in Bauer et al. (2015) is L$_{\rm X}=2.2 \times 10^{43}$ erg s$^{-1}$ and indeed, the authors state that it is a factor $\sim$1.6 lower than the one derived from the mid-IR to X-ray relation in Gandhi et al. (2009) (which is the orange vertical stripe in Fig. 3). However, we note that at high column densities the derived intrinsic luminosity is highly dependent on the (unknown) geometry of the absorber \citep[e.g.][]{mpl99}.
The fit does not improve if we leave the normalization and column density of the 2015 spectra free to vary, indicating that there is no difference between the 2012 and 2015 data sets. 
Residuals around 25-30 keV in both August 2014 and February 2015 observations (Fig. 2, middle and bottom panels) may be ascribed to residual instrumental features in the NuSTAR ARFs (see Fig. 7 and 8, Madsen et al. 2015). If we include the instrumental background emission lines between 22 keV and 35 keV, no significant variations in the best fit parameters are found. This effect represents $\sim2\%$ of the total 20-80 keV flux in the August 2014 observation and $\sim7\%$ of the observed flux excess, in the 20-80 keV energy band.
Fig. \ref{cp} shows the contour plots between the column density along the line of sight (the one directly obscuring the primary continuum) and the intrinsic 2-10 keV nuclear luminosity extrapolated from the de-absorbed best fit from the primary continuum for the three {\it NuSTAR} data sets (colors as in Fig. \ref{PLOT}). The inferred X-ray luminosity L$_{\rm X}\sim 10^{43}$--$10^{44}$ erg s$^{-1}$ is almost four orders of magnitude greater than usually observed in ULXs \citep{sgt04, wrm11}:  the lack of variations below 10 keV and the sharp cutoff in ULX spectra above 20 keV \citep{brw13, wfh13, whg14, wmr15} lead us to conclude that this transient excess cannot be attributed to the ULX in the NGC 1068 FOV.

We then considered the possibility that the high-energy excess in the August 2014 observation might be due to a rise in the reflected emission only, due to matter with N$_{\rm H}\sim 10^{25}$ cm$^{-2}$. The only way to have a variation in the Compton hump without an associated variation in the iron line is for the reflector
to be almost completely self-obscured. Indeed, fits show that the inclination angle of the reflector would have to be larger than 87$^{\circ}$, assuming a toroidal configuration. 
We therefore conclude that this intepretation, while not impossible, is unlikely. 
\section{Discussion}
We interpret the high-energy excess detected in the August 2014  {\it NuSTAR} spectra as the first unveiling event ever observed in NGC 1068, in which there is a drop in the column density along our line of sight. If we take into account the mid-IR and [OIII] luminosities as proxies of the intrinsic nuclear luminosity, we have additional pieces of information to add to the contour plots shown in Fig. \ref{cp}.  The vertical orange lines represent the intrinsic 2-10 keV luminosity inferred from the mid-IR luminosity \citep{ghs09}. Purple lines indicate the intrinsic 2-10 keV luminosity calculated from the extinction-corrected [OIII] luminosity \citep{mbn12} using the [OIII]-X ray relation from \citet{lamastra09}. Contour  plots show that the intrinsic X-ray luminosity for the three observations is consistent with those inferred using other proxies, and all the spectral difference can be attributed to a change in the absorbing column density, from N$_{\rm H}>8.5\times 10^{24}$ cm$^{-2}$ in the 2012 observation to N$_{\rm H,1}=(5.9\pm0.4)\times 10^{24}$ cm$^{-2}$ in 2014 (Fig. \ref{cp}, using 90\% confidence level for two interesting parameters).

Assuming the bolometric correction from \citet{mar04} we infer L$_{\rm bol}=2.1^{+3.2}_{-1.4}\times 10^{45}$ erg s$^{-1}$, in agreement with \citet{hpb08}. The black hole mass of NGC 1068 is estimated to be $\simeq1\times 10^7$ M$_{\odot}$ \citep{gga96, lb03}. For consistency with the mid-IR and [OIII] luminosities (Fig. \ref{cp}) we take the lower value L$_{\rm bol}=7\times 10^{44}$ erg s$^{-1}$, leading to an accretion rate $\eta_{\rm edd}\sim0.55$, confirming the highly accreting nature of the source.

Absorption variability is common when observations performed months to years apart are compared \citep{risa02b}, and has been found on time scales of hours to days in several sources. However, even in the so-called ``changing-look AGN'' (sources that switched from the Compton-thick to the Compton-thin state and vice versa) an eclipsing/unveiling event affecting only the spectrum above 10 kev has never been observed: we emphasize that this is the first time that a Compton-thick unveiling event of this kind has been reported. We note that this is different from the intrinsic variability recently reported for the Compton-thick AGN NGC 4945 \citep{pcf14}. Our finding is supporting a clumpy structure of the obscuring material along the line of sight \citep{nsn08}. 

In this scenario we do not have a single, monolithic obscuring wall, but the total column density along the line of sight is the sum of the contributions from a discrete number of clouds. The {\it NuSTAR} sensitivity above 10 keV allowed us to infer only a lower limit on the column density variation ($\Delta$N$_{\rm H}\simeq2.5\times 10^{24}$ cm$^{-2}$) but greater changes could have occurred (top-left corner of Fig. 3: the parameter space with N$_{\rm H}>8.5\times 10^{24}$ cm$^{-2}$) but were not measurable with our data. Further monitoring of NGC 1068 could provide contraints on the number of clouds and their distance from the illuminating source.

\section{Conclusions}
We presented a spectral analysis of the 4-79 keV {\it NuSTAR} and XMM-{\it Newton} monitoring campaign of NGC 1068 obtained between July 2014 and February 2015. We found a clear transient excess above 20 keV in the August 2014 {\it NuSTAR} observation, while no variations are found in the XMM data below 10 keV.
The most plausible explanation is an unveiling event, in which for a short while the total absorbing column, probably composed by a number of individual clouds, became less thick so as to permit to the nuclear radiation to pierce through it. Our result provides further evidence that the obscuring material along our line of sight is clumpy, and enables us to infer a 2--10 keV intrinsic luminosity of L$_{\rm X}=7^{+7}_{-4} \times 10^{43}$ erg s$^{-1}$.

\section*{ACKNOWLEDGEMENTS}
We thank the referee for her/his comments. AM, SB and GM acknowledge financial support from Italian Space Agency under grant ASI/INAF I/037/12/0-011/13. FEB acknowledges support from CONICYT-Chile (PFB-06/2007, FONDECYT 1141218, ACT1101), and grant IC120009, awarded to The Millennium Institute of Astrophysics, MAS. WNB acknowledges Caltech NuSTAR subcontract 44A-1092750. This work was supported under NASA Contracts No. NNG08FD60C, NNX10AC99G, NNX14AQ07H and made use of data from the {\it NuSTAR} mission, a project led by the California Institute of Technology, managed by the Jet Propulsion Laboratory, and funded by the National Aeronautics and Space Administration. We thank the {\it NuSTAR} Operations, Software and Calibration teams for support with the execution and analysis of these observations.  This research has made use of the {\it NuSTAR} Data Analysis Software (NuSTARDAS) jointly developed by the ASI Science Data Center (ASDC, Italy) and the California Institute of Technology (USA).

\bibliographystyle{mn2e}
\bibliographystyle{mn2e}
\bibliography{sbs} 

\end{document}